\documentclass[%
 reprint,
superscriptaddress,
showpacs,
 amsmath,amssymb,
 aps,
pra,
]{revtex4-1}

\usepackage{graphicx}
\usepackage{tabularx}
\usepackage{amsmath}
\usepackage{float}

\newcommand{\unit}[1]{\,\mbox{#1}}
\newcommand{\um}{\unit{$\mu$m}}
\newcommand{\us}{\unit{$\mu$s}}
\newcommand{\G}{\unit{G}}
\newcommand{\kHz}{\unit{kHz}}
\newcommand{\MHz}{\unit{MHz}}
\newcommand{\degree}{\mbox{$^{\circ}$}}
\newcommand{\degC}{\mbox{\degree{}C}}

\newcommand{\ish}{\mbox{$\sim$}\,}
\newcommand{\bra}[1]{\mbox{$\left< #1 \right|$}}
\newcommand{\ket}[1]{\mbox{$\left| #1 \right>$}}

\newcommand{\ion}[2]{\mbox{$^{#2}$#1$^+$}}
\newcommand{\Yb}[1]{\ion{Yb}{#1}}

\begin{document}
\preprint{APS/123-QED}

%\title{Production of Schrodinger cat states using microwaves within a static magnetic field gradient.}
\title{Generation of spin-motion entanglement in a trapped ion using long-wavelength radiation}

\author{K. Lake}
\affiliation{Department of Physics and Astronomy, University of Sussex, Brighton, BN1 9QH, UK}
\author{S. Weidt}
\affiliation{Department of Physics and Astronomy, University of Sussex, Brighton, BN1 9QH, UK}
\author{J. Randall}
\affiliation{Department of Physics and Astronomy, University of Sussex, Brighton, BN1 9QH, UK}
\affiliation{QOLS, Blackett Laboratory, Imperial College London, London, SW7 2BW, UK}
\author{E. Standing}
\affiliation{Department of Physics and Astronomy, University of Sussex, Brighton, BN1 9QH, UK}
\author{S. C. Webster}
\affiliation{Department of Physics and Astronomy, University of Sussex, Brighton, BN1 9QH, UK}
\author{W. K. Hensinger}
\email[]{w.k.hensinger@sussex.ac.uk}
\affiliation{Department of Physics and Astronomy, University of Sussex, Brighton, BN1 9QH, UK}

%\date{\today}

\begin{abstract}
Applying a magnetic field gradient to a trapped ion allows long-wavelength radiation to produce a mechanical force on the ion's motion when internal transitions are driven. We demonstrate such a coupling using a single trapped  \Yb{171}~ion, and use it to produce entanglement between the spin and motional state, an essential step towards using such a field gradient to implement multi-qubit operations.
\end{abstract}

\maketitle

\section{Introduction}
Systems of trapped ions have proven to be an ideal physical system for demonstrating small-scale quantum information processing and quantum simulation \cite{09:home, 10:kim,11:lanyon, haffner, 12:khromova}. By making use of the strong motional coupling between co-trapped ions, numerous different implementations of high-fidelity quantum gates between individual ions have been demonstrated \cite{00:sackett,03:leibfriedb,03:schmidtb,06:home,08:benhelm}. Many of these implementations make use of spin dependent forces which allow the use of ions in thermal motion. 

Producing quantum devices that match or exceed the capabilities offered by current `classical' computers will require scaling the systems to much larger numbers of ions, with large numbers of gate operations performed in parallel. For gates based on laser radiation this will require laser light with high powers that has minimal intensity fluctuations as well as intricate optical control to deliver the light to the correct ions without cross-talk \cite{07:leibfried}. One possible alternative to this is to instead make use of microwave or radio-frequency (RF) radiation to drive gates since the generation of intense stable radiation at these frequencies is much easier. Such long-wavelength radiation by itself is, however, not able to generate the forces required to implement these multi-ion gates.

A magnetic field gradient at the position of the ions can create a state dependent force which increases the coupling of the radiation to the ion's motional state \cite{01:mintert} and would allow a broad range of gates previously implemented using laser light to instead be driven by long-wavelength radiation. Such a field gradient also allows for individual addressing of trapped ions by a global radiation field with extremely small cross-talk \cite{09:johanning, 14:piltz}, as well as producing an intrinsic spin-spin coupling between ions even when no microwave or RF radiation is applied, which can be used for gate operations  \cite{12:khromova}. Another approach is to place the ions in the near-field oscillating magnetic field gradient of a microwave waveguide or waveguide cavity \cite{08:ospelkaus, 11:ospelkaus} to enable the coupling to the motion allowing for multi-ion gates.

\begin{figure}[tb]
\centering
\includegraphics[width=0.49\textwidth]{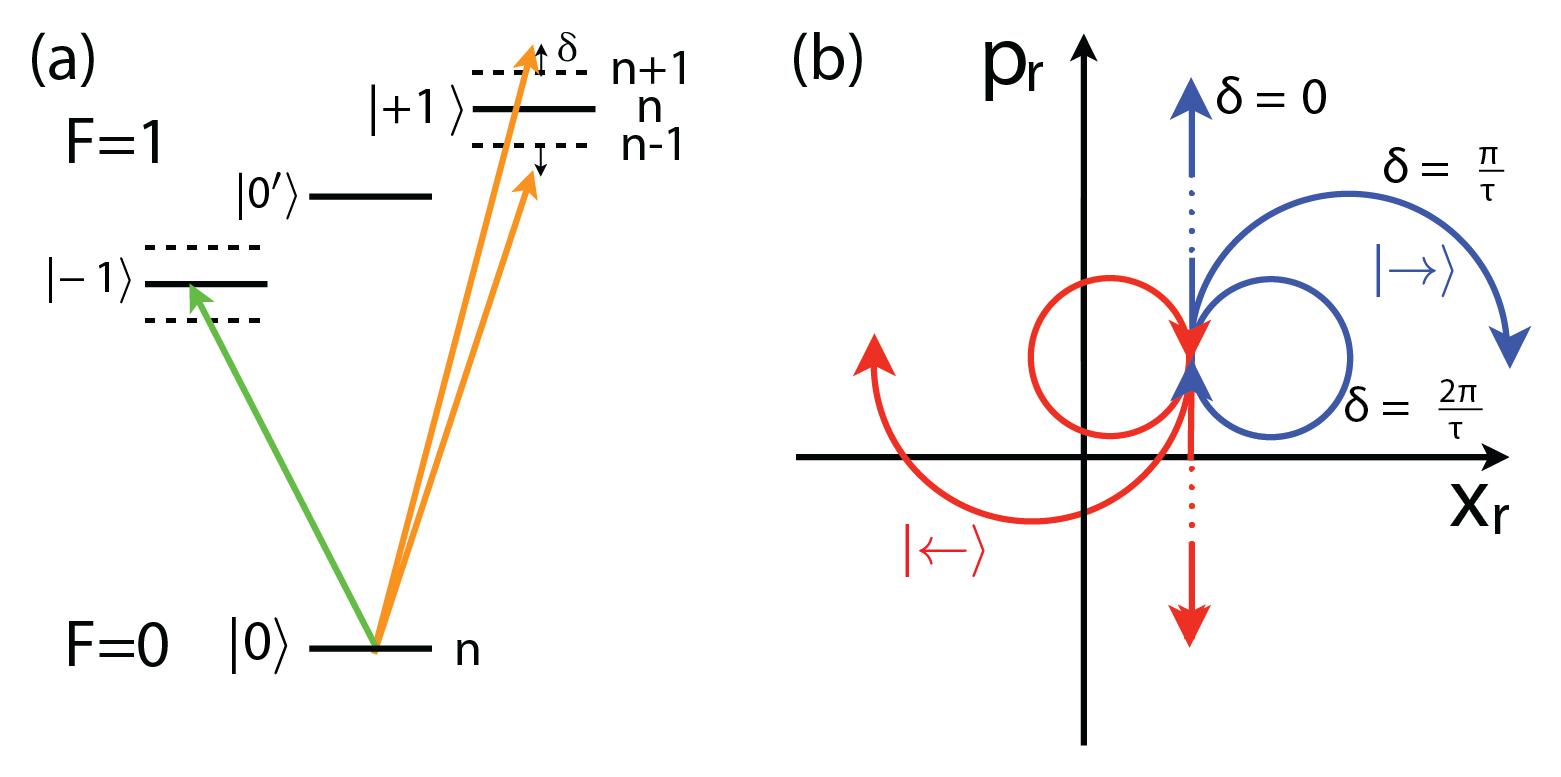}
\caption{(a) Energy level diagram of the ground state of \Yb{171} consisting of the $F=0$ state \ket{0} and three $F=1$ states \ket{-1}, \ket{0'} and \ket{+1}. Due the presence of a magnetic field gradient a single microwave field (indicated in green) can drive motional state changing transitions from \ket{0}~to either \ket{-1}~or \ket{+1}. A pair of microwave fields with equal but opposite detunings $\delta$ from the two motional sidebands (shown in orange) produces a spin-dependent M{\o}lmer-S{\o}rensen force on the ion. (b) The M{\o}lmer-S{\o}rensen force displaces the two orthogonal spin states (both equal superpositions of \ket{0}~and in this case \ket{+1} with phases related to the phases of the driving fields) in opposite directions in motional phase space. The displacements describe circular arcs in phase space, with one complete circle obtained for a driving time $\tau=2\pi/\delta$.}
\label{fig:levels}
\end{figure}

Here we report on the use of microwave radiation in conjunction with a large static magnetic field gradient to generate a coupling between the internal and motional states of a trapped ion and we demonstrate spin-motion entanglement using this method. This coupling could then be used to implement multi-qubit quantum gates such as the M{\o}lmer-S{\o}rensen gate \cite{00:sorensen, 11:timoney}. We demonstrate this coupling by using it to produce entanglement between the internal spin state of a single trapped ion and its external motional state (a Schr\"odinger cat state), a single-ion version of the multi-ion M{\o}lmer-S{\o}rensen gate \cite{96:monroe, 05:haljan}.

A measure of the coupling strength of radiation to a motional mode of a trapped ion is given by the Lamb-Dicke parameter $\eta=k_z z_0$, where $k_z$ is the projection of the effective wave vector along the mode direction and $z_0=\sqrt{\hbar/2m\nu_z}$ is the spatial extent of the ground-state wave function of the mode. The effective wave vector is the wave vector $\mathbf{k}$ of the coupling radiation field for single photon transitions, and the difference wave vector $\mathbf{\delta k}=\mathbf{k}_1-\mathbf{k}_2$ for stimulated Raman transitions. $m$ is the mass of the ion and $\nu_z$ the frequency of the motional mode. This is negligibly small for transitions driven between hyperfine states of the ground state of \Yb{171}~which lie in the microwave regime, however applying a static magnetic field gradient to the ion produces an effective Lamb-Dicke parameter $\eta_{\rm eff}=z_0 \Delta F/\hbar\nu_z$ where $\Delta F$ is the differential force experienced by the two ionic states involved in the transition \cite{01:mintert}.

For the transition \ket{0}$\leftrightarrow$\ket{+1} (see figure \ref{fig:levels} (a) for state definitions) and an ion experiencing a magnetic field $B=B_0+z\partial_zB$, $\Delta F=\mu_B\partial_zB$, resulting in an effective Lamb-Dicke parameter of 
\begin{equation}
\eta_{\rm eff} = \frac{z_0\mu_B\partial_zB}{\hbar\nu_z}=\frac{\mu_B\partial_zB}{\sqrt{2m\hbar}\nu_{z}^{3/2}}.
\end{equation}

\begin{figure}[tb]
\centering
\includegraphics[width=0.45\textwidth]{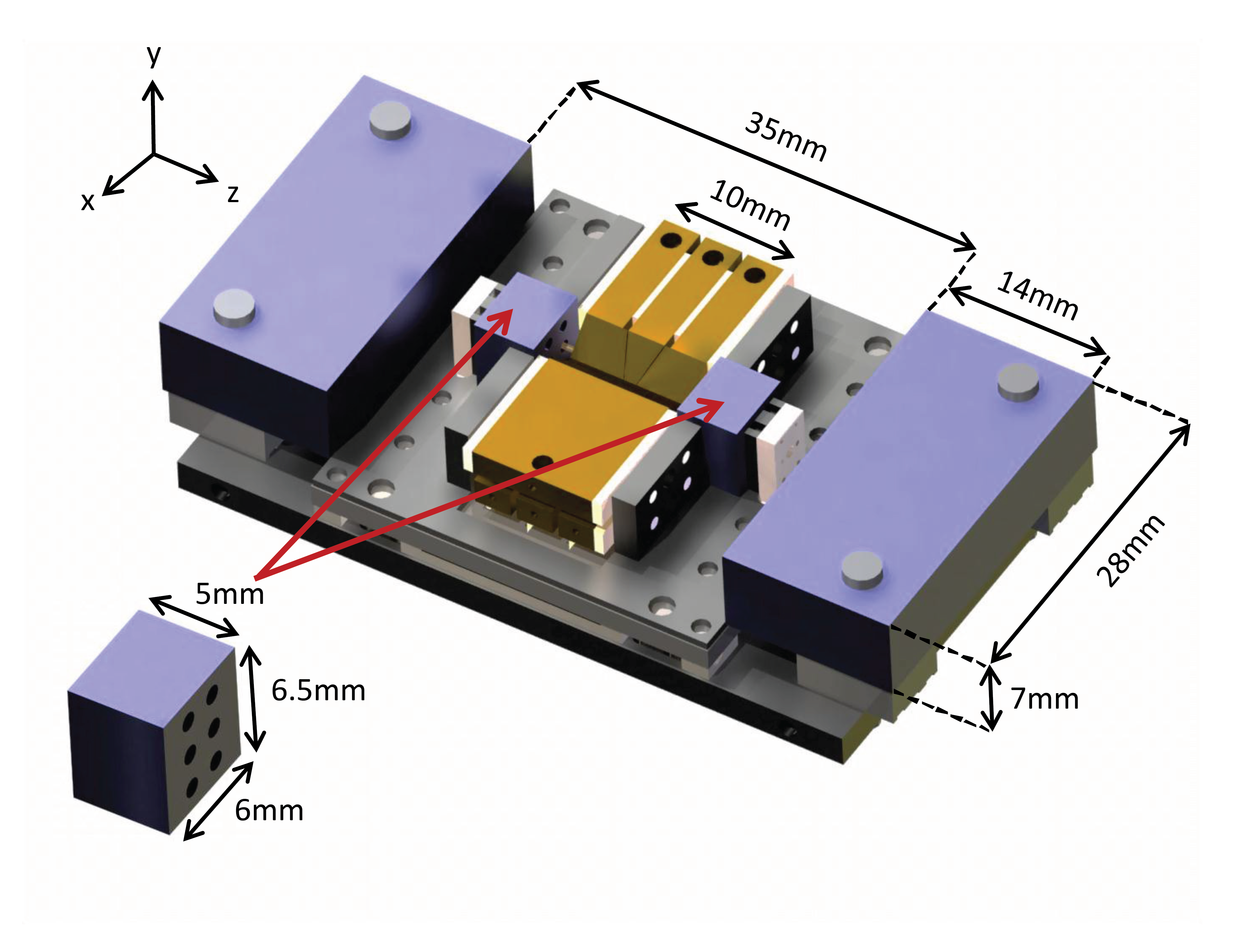}
\caption{Schematic diagram showing the integration of the magnets into the trap structure. The four magnets are coloured blue, and the trap electrodes golden. The smaller pair of magnets have holes machined in them to allow the trap's compensation electrodes to pass through them. See  \cite{11:mcloughlin} for a full description of the trap structure.}
\label{fig:magnets}
\end{figure}

\section{creating a large magnetic field gradient}
To create a magnetic field gradient along the trap axis four permanent magnets are integrated into the structure of the trap electrodes, as shown in figure \ref{fig:magnets}. The magnets are made of Samarium Cobalt, which combines a high residual magnetic flux density with a high Curie temperature of 800\degC. This minimises possible demagnetisation of the magnets during the baking of the vacuum system. The magnets are coated in nickel copper nickel to ensure they are UHV compatible, and machined to a nominal precision of 50\um. While the magnets are designed to produce a null at the trap centre, imperfections in magnet placement requires offset fields of \ish100\G~to be applied by external field coils to leave a small bias field at the ion's position.

\begin{figure}[tb]
\centering
\includegraphics[width=0.45\textwidth]{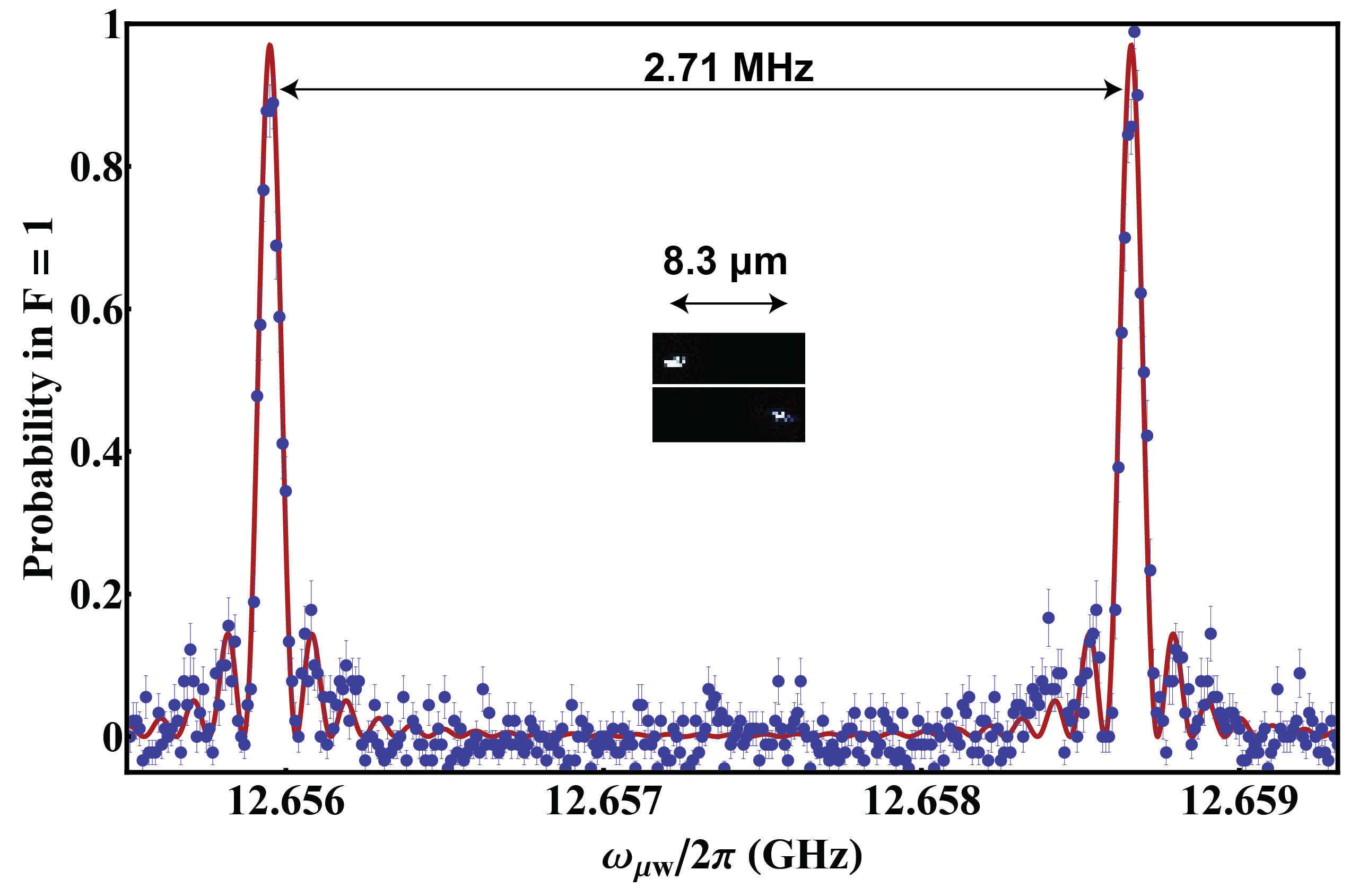}
\caption{The probability to drive the $\ket{0} \leftrightarrow \ket{-1}$ transition for two ions trapped in a magnetic field gradient. The gradient causes the transition frequencies to be different, allowing for individual addressing in frequency space. The red line is a theory curve describing the sum of two transition probabilities, with the two resonant transition frequencies and the Rabi frequency as free parameters. The axial secular frequency $\nu_z$ was measured to be $2\pi \times 268$ kHz, implying an axial magnetic field gradient of $23.3(6)$ Tm$^{-1}$. Each point is an average of 200 measurements.}
\label{fig:indadd}
\end{figure}

The magnetic field gradient was determined by measuring the $\ket{0}\leftrightarrow\ket{-1}$ transition frequencies of a pair of trapped ions. Both ions are prepared in F=0, and after the application of microwave radiation equivalent to a $\pi$ pulse it is determined if an ion has transitioned to F=1. A description of the methods used for preparation and measurement of the hyperfine state is given in \cite{13:webster}. Figure \ref{fig:indadd} shows the resultant probability of at least one ion being transferred to F=1 as a function of the frequency of the microwave pulse; the two peaks separated by $\Delta\omega = 2\pi\times 2.71\MHz$~correspond to the two different transition frequencies of the different ions. In order to accurately determine the spatial separation between the two ions we measure the axial centre of mass frequency $\nu_z$ by performing motional sideband spectroscopy (as described in the next section). From this we measure $\nu_z=2\pi\times268\,{\rm kHz}$ and calculate an ion-ion separation of 8.3\um, giving an axial magnetic field gradient $\partial_zB$ of $23.3(6)\,{\rm Tm}^{-1}$.

Any drift in the magnetic field or ion positions during the course of the scan will lead to an error in the  gradient measurement. By interleaving the data taking for the measurements of the two resonant frequencies they can be measured near simultaneously, eliminating the effect of drifts. This measurement gives a more accurate gradient measurement of  $23.6(3)\,{\rm Tm}^{-1}$.

This figure also illustrates the near perfect individual addressing of co-trapped ions illuminated by a global radiation field. The non-resonant excitation probability is less than $(\Omega/\Delta\omega)^2$ \cite{12:khromova} and for a Rabi frequency of $2\pi\times 40\kHz$ this corresponds to a probability of less than $2\times 10^{-4}$.

\section{spin-motion entanglement}
For the $\ket{0}\leftrightarrow\ket{+1}$ transition in $^{171}$Yb$^+$, trapped with an axial motional frequency $\nu_z=2\pi\times 268\kHz$, the effective Lamb-Dicke parameter resulting from the magnetic field gradient is $\eta_{\rm eff}=0.013$. While this is smaller than Lamb-Dicke parameters generated by laser fields (a pair of counter propagating Raman beams near the $S_{1/2}\leftrightarrow P_{1/2}$ transition give $\eta=0.36$ for this trap frequency), it can be used to produce a two-qubit gate \cite{00:sorensen}.

\begin{figure}[tb]
\centering
\includegraphics[width=0.45\textwidth]{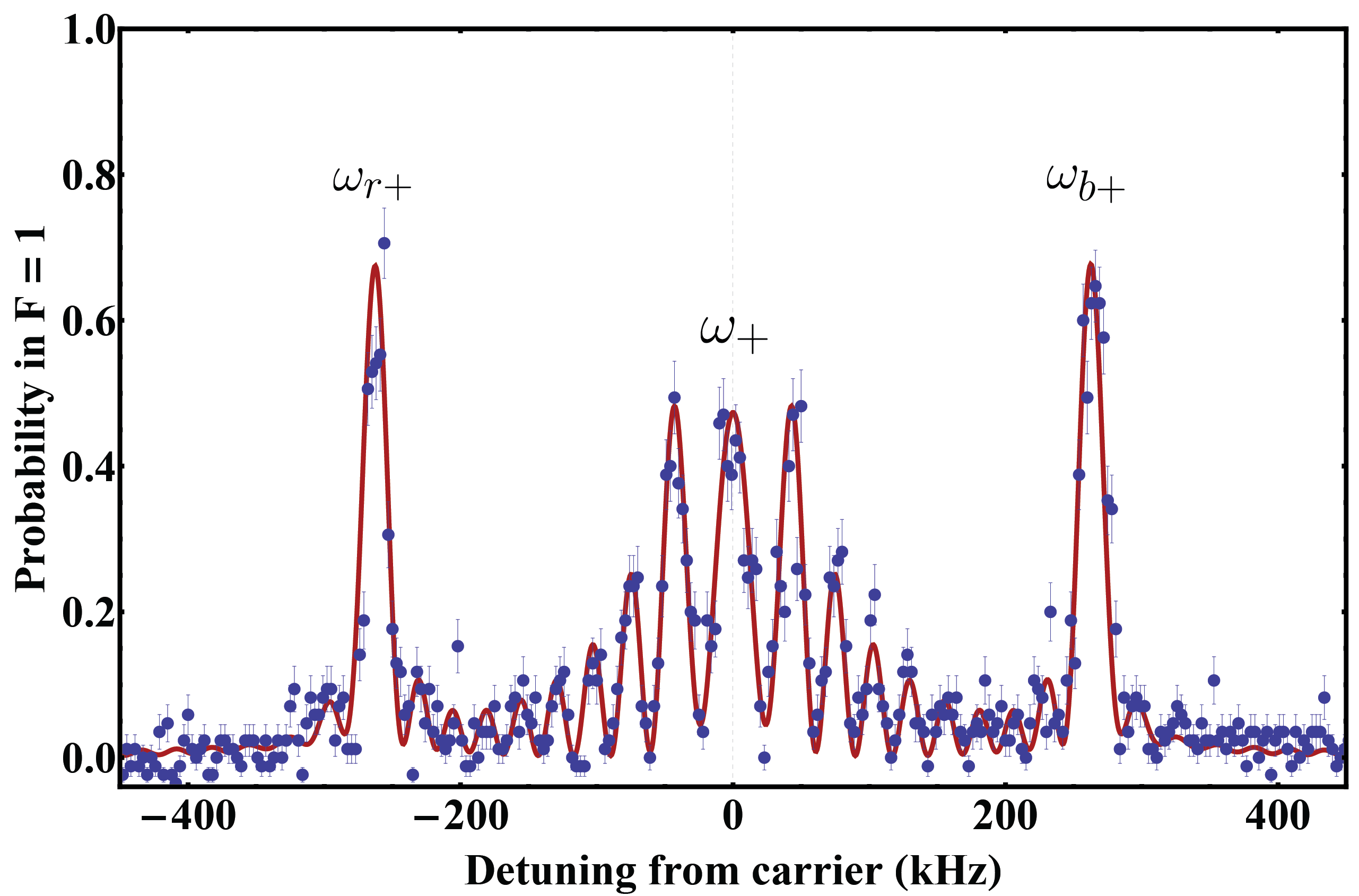}\\
\caption{Probability of exciting an ion to F=1 as a function of frequency for a 40\us~microwave pulse, clearly showing the resolved sideband structure of the transition. The sideband peaks are separated in frequency from the carrier by the secular frequency, $\nu_z / 2 \pi = 268$\,kHz. The carrier Rabi frequency is $2 \pi \times 46$\,kHz. Each point is an average of 200 measurements. The red line is a theory curve classically summing up the responses on the carrier and two sidebands with a carrier Rabi frequency of $2 \pi \times 46$\,kHz, a secular frequency of $\nu_z / 2 \pi = 268$\,kHz and with $\overline{n}$ as a free parameter. The thermal distribution of phonon states determined from this curve is $\overline{n}=290(50)$.}
\label{fig:motionalsb}
\end{figure}

Figure \ref{fig:motionalsb} demonstrates the existence of this motional coupling, by showing the presence of sidebands separated from the carrier transition frequency by the motional trap frequency. From the theory curve in figure \ref{fig:motionalsb} a mean motional excitation following Doppler cooling of $\overline{n}=290(50)$ has been determined, assuming a thermal distribution. While this is far from the ground state, the ion is still well within the Lamb-Dicke regime.

By applying a pair of microwave fields to the ion, one detuned from the blue sideband and the other with equal but opposite detuning from the red sideband, a state-dependent force can be produced. Driving an ion with this force can result in the production of a Schr\"odinger cat state, where the ion's spin is entangled with its motional state \cite{96:monroe, 05:haljan}. This state dependent force is the same as is applied to two or more ions to perform a M{\o}lmer-S{\o}rensen entangling gate \cite{99:molmer, 00:sorensen}, and its application to a single ion is an important precursor to being able to perform such a gate driven by microwave radiation. The M{\o}lmer-S{\o}rensen gate only requires that the ions are in the Lamb-Dicke regime, making it a powerful tool for quantum logic with trapped ions.

The Hamiltonian describing our system where two microwave fields of equal Rabi frequency $\Omega$ with frequencies $\omega_++\nu_z-\delta$ and $\omega_+-\nu_z+\delta$ are applied to the ion, after making the rotating wave approximation and moving to the interaction picture, can be written as 
\begin{equation}
\hat{H}_{I}= \frac{\hbar \eta_{\rm eff} \Omega}{2}(\sigma_++\sigma_-) (\hat{a}^{\dagger} e^{-i\delta t} + \hat{a} e^{i\delta t}) 
\label{eq:sigman234}
\end{equation}
where for simplicity we have set the phases of the microwave fields to zero. Such a Hamiltonian causes a spin dependent displacement of the motional state, specifically on the spin states $\ket{\rightarrow} = (\ket{0}+\ket{+1})/\sqrt{2}$ and $\ket{\leftarrow} = (\ket{0}-\ket{+1})/\sqrt{2}$, as shown on the phase space diagram in figure \ref{fig:levels}b. Different choices for the microwave phases would result in different phases of these spin superpositions. The evolution of the system is described by the unitary operator 
\begin{equation}
U(t)=D(\alpha(t))\ket{\rightarrow}\bra{\rightarrow} + D(-\alpha(t))\ket{\leftarrow}\bra{\leftarrow}.
\end{equation}
$D({\alpha})$ is a motional displacement operator, $\alpha(t)=\alpha_0(1-e^{-i\delta t})$ describes a circular motion of the ion's displacement in phase space with $\alpha_0=\eta_{\rm eff}\Omega/2\delta$ and where we ignore a global phase caused by the displacement \cite{footnote, 05:haljan}.

An ion initially in the state \ket{0} can be written in the M{\o}lmer-S{\o}rensen spin basis as the superposition $(\ket{\rightarrow}+\ket{\leftarrow})/\sqrt{2}$. Under the application of the M{\o}lmer-S{\o}rensen force the two spin states undergo different motional displacements, causing entanglement between the ion's spin and motional states. These different displacements mean that the two spin-states can in principle be distinguished, and so if the ion's spin is measured the superposition of spin states is dephased, reducing the probability of measuring the ion to still be in \ket{0}, with the degree of dephasing depending on the distinguishability of the two motional states. The spin is least sensitive to such displacement-induced depolarisation if initially prepared in the $n=0$ ground state (i.e. the coherent state \ket{\alpha=0}), in which case the state evolves as

\begin{equation}
\ket{\psi(t)} = \frac{1}{\sqrt{2}}\bigg(\ket{\rightarrow}\ket{\alpha(t)} + \ket{\leftarrow}\ket{-\alpha(t)}\bigg).
\end{equation}
The probability for the ion to be in the state $\ket{+1}$ is then given by

\begin{equation}\begin{split}
P_{\left|+1\right>} &= \big|\left<+1|\psi(t)\right>\big|^2 \\
&= \frac{1}{2}-\frac{1}{2}\exp\left(-2\left|\frac{\eta_{\rm eff}\Omega}{2\delta}\left(1-e^{-i\delta t}\right)\right|^2\right).
 \label{eq:probup}
\end{split}\end{equation}
For increasing energy of the initial motional state, the sensitivity of the state \ket{0} to depolarisation increases, such that if the ion begins in a thermal state characterised by mean vibrational occupation number $\overline{n}$, Eq. \ref{eq:probup} is modified to read \cite{05:haljan}
\begin{equation}
P_{\left|+1\right>}=\frac{1}{2}-\frac{1}{2}\exp\left(-2\left(2\overline{n}+1\right)\left|\frac{\eta_{\rm eff}\Omega}{2\delta}\left(1-e^{-i\delta t}\right)\right|^2\right)
\label{eq:probupnbar}
\end{equation}

\begin{figure}[H]
\centering
\includegraphics[width=0.45\textwidth]{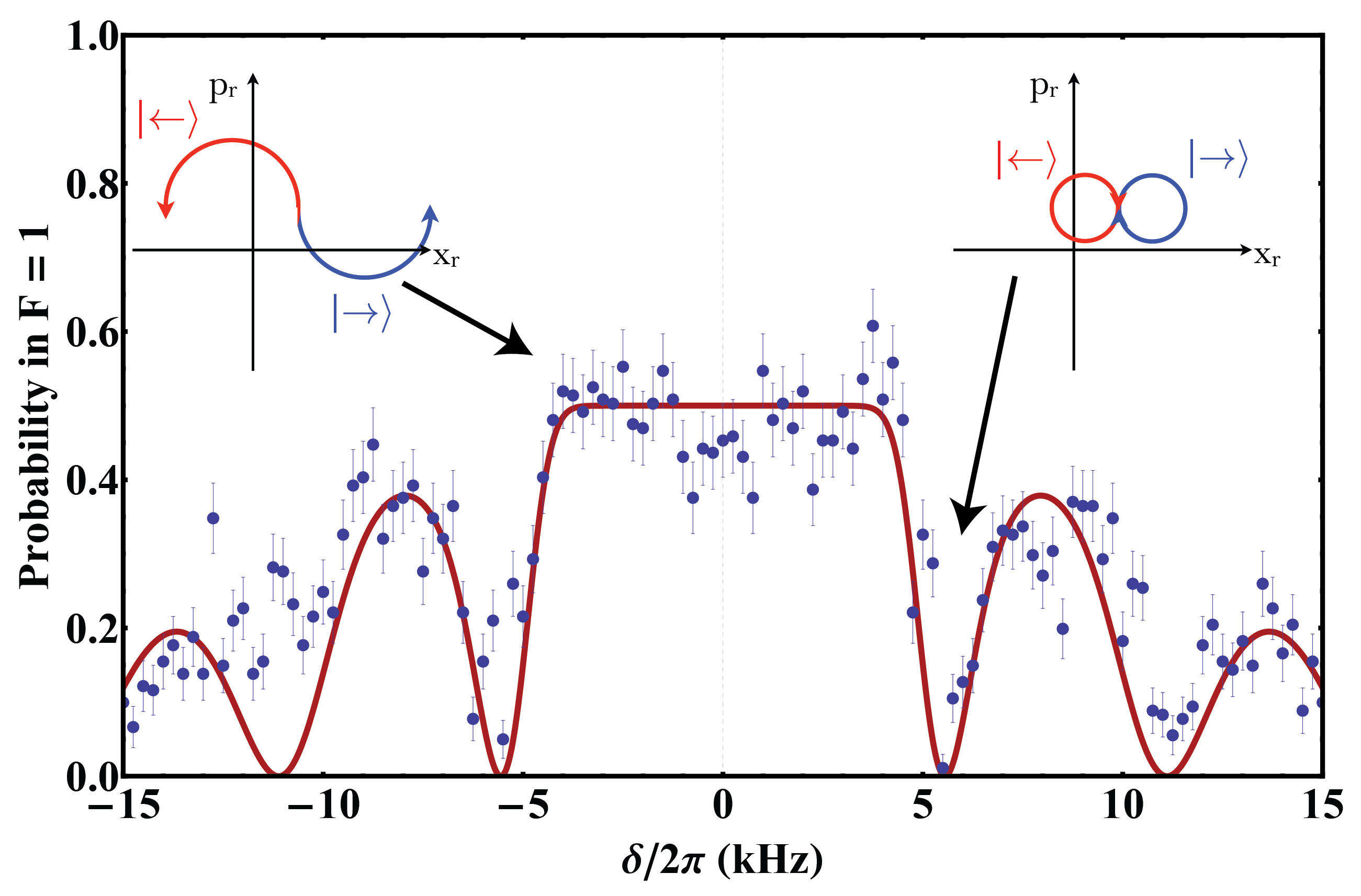}
\caption{Probability of an ion initially in the \ket{0} state being in $F=1$ after the application of a M{\o}lmer-S{\o}rensen force to the $\ket{0}\leftrightarrow\ket{+1}$ transition for a fixed time $\tau=180\us$. The carrier Rabi frequency $\Omega$ was  $2 \pi \times 35$\kHz. Each point is an average of 200 measurements. A theory line is also shown for an initial thermal excitation $\overline{n}=110$. The data deviates from the theory line for a detuning of -11\kHz~due to slow drifts in the applied magnetic field that were not compensated for when taking the data. The insets show the displacement of the orthogonal spin states in motional phase space for different detunings.}
\label{fig:cat}
\end{figure}

Figure \ref{fig:cat} shows the result of applying the M{\o}lmer-S{\o}rensen force to a single \Yb{171}~ion with variable detuning $\delta$ for a time $\tau=180\us$, before measuring the spin state. The amount of depolarisation of the initial \ket{0} state varies with detuning, with the depolarisation periodically reducing close to zero, at detunings $\delta_j=2\pi j/\tau$ for non-zero integers $j$. These detunings correspond to the zero resultant microwave-induced displacements after the application of the pulses, the spin states having undergone $j$ complete loops in phase space resulting in no residual entanglement between spin and motion. The maximum distance between the two motional states in phase space occurs when $\delta=0$. For the parameters used in Fig. \ref{fig:cat}, the maximum distance is $\Delta\alpha_{\rm max}=0.28$.

The solid line in figure \ref{fig:cat} is the result of applying equation \ref{eq:probupnbar} with the known carrier Rabi frequency $\Omega/2\pi=35\kHz$ and an initial thermal excitation of $\overline{n}=110$ (the cooling parameters were further optimised compared with the data shown in figure 4 resulting in a significantly lower $\overline{n}$). Since $\overline{n}\gg1$, only a small displacement in phase space between the two spin states is required to produce complete depolarisation. Such a spin-dependent force is essential to being able to implement a M{\o}lmer-S{\o}rensen entangling gate.

\section{conclusion}
We have described the creation of a large magnetic field gradient in the centre of an ion trap, and demonstrate that it can produce a significant coupling between an ion's spin state and its motion when driven by microwave radiation. We then use this coupling to create a spin-dependent force to generate entanglement between spin and motion, an important step to being able to produce a M{\o}lmer-S{\o}rensen gate between two ions using microwave radiation.

This coupling requires a transition which is very sensitive to a magnetic field to be realised, which can result in low decoherence times due to environmental fluctuations in the field. By dressing the ion with microwave fields, the decoherence time of the qubit can be massively increased, while still retaining the coupling due to the field gradient \cite{11:timoney, 13:webster}. This will allow multi-qubit quantum gates to be realised using microwave fields.

\section{Acknowledgements}

We would like to thank Diego Porras for useful discussions.

This work is supported by the UK Engineering and Physical Sciences Research Council (EP/E011136/1, EP/G007276/1), the European Commission's Seventh Framework Programme (FP7/2007-2013) under grant agreement no. 270843 (iQIT), the Army Research Laboratory under Cooperative Agreement Number W911NF-12-2-0072 and the University of Sussex. The views and conclusions contained in this document are those of the authors and should not be interpreted as representing the official policies, either expressed or implied, of the Army Research Laboratory or the U.S. Government. The U.S. Government is authorized to reproduce and distribute reprints for Government purposes notwithstanding any copyright notation herein.

\bibliography{paperrefs}

\begin{thebibliography}{24}
\bibitem [{foo()}]{footnote}%
  \BibitemOpen
  \href@noop {} {}\bibinfo {note} {Note however that this global phase is the
  cause of the entangling effect when such a force is applied to two ions to
  drive a M{\o}lmer-S{\o}rensen gate}\BibitemShut {NoStop}%
\end{thebibliography}

\end{document}